\documentclass[onecolumn]{article}
\usepackage{amsmath,amssymb,caption,graphicx,epstopdf,mathtools,color,cite,url}
\usepackage{physics}

\DeclareMathOperator{\erfc}{erfc}

\newcommand{\erfceqdim}{\erfc\Big(\sqrt{\dfrac{m}{2\overline{T}}}\dot{\overline{x}}\Big)}
\newcommand{\expeqdim}{\exp\Big(-\frac{m \dot{\overline{x}}^2}{2\overline{T}}\Big)}
\newcommand{\erfceqadim}{\erfc\Big( \frac{\epsilon\dot{x}}{\sqrt{2T}}\Big)}
\newcommand{\expeqadim}{\exp\Big(-\frac{\epsilon^2\dot{x}^2}{2T}  \Big)}

\newcommand{\ndot}[1]{\stackrel{\scriptstyle{(n)}}{#1}}
\newcommand{\deq}{\text{d.eq.}}
\newcommand{\app}{\text{ap}}
\newcommand{\half}{\frac{1}{2}}

\renewcommand{\epsilon}{\varepsilon}
\newcommand{\ud}{\mathrm{d}}
\newcommand{\btau}{\boldsymbol{\tau}}

\makeatletter 
\renewcommand{\pdv}[2]{\begingroup 
  \@tempswafalse\toks@={}\count@=\z@ 
  \@for\next:=#2\do 
    {\expandafter\check@var\next\@nil
     \advance\count@\der@exp 
     \if@tempswa 
       \toks@=\expandafter{\the\toks@\,}% 
     \else 
       \@tempswatrue 
     \fi 
     \toks@=\expandafter{\the\expandafter\toks@\expandafter\partial\der@var}}% 
  \frac{\partial\ifnum\count@=\@ne\else^{\number\count@}\fi#1}{\the\toks@}% 
  \endgroup} 
\def\check@var{\@ifstar{\mult@var}{\one@var}} 
\def\mult@var#1#2\@nil{\def\der@var{#2^{#1}}\def\der@exp{#1}} 
\def\one@var#1\@nil{\def\der@var{#1}\chardef\der@exp\@ne} 
\makeatother 

\newcommand{\BigO}[1]{\mathcal{O}\left( #1 \right)}
\newcommand{\varepsilont}{\varepsilon t}
\newcommand{\epsilont}{\varepsilont}
\newcommand{\IF}{\int_{0}^{t}F\left( \varepsilon \chi \right)\ud\chi}

\allowdisplaybreaks

\begin{document}

\title{Multiple scales approach to the Gas-Piston non-equilibrium themodynamics}
\author{D. Chiuchi\`u$^1$ \and G. Gubbiotti$^2$}

\date{$^1$ NiPS Lab, Universit\'a degli Studi di Perugia, Dipartimento di Fisica e Geologia \footnote{\texttt{davide.chiuchiu@nipslab.org}}\\ $^2$ Universit\'a degli Studi Roma tre, Dipartimento di Matematica e Fisica and Sezione INFN di Roma Tre \footnote{\texttt{gubbiotti@mat.uniroma3.it}}}

\maketitle

\begin{abstract}
The non-equilibrium thermodynamics of a gas inside a piston is a conceptually simple problem where analytic results are rare. For example, it is hard to find in the literature analytic formulas that describe the heat exchanged with the reservoir when the system either relaxes to equilibrium or is compressed over a finite time. In this paper we derive such kind of analytic formulas. To achieve this result, we take the equations derived by Cerino \textit{et al.} [Phys. Rev. E \textbf{91}, 032128] describing the dynamic evolution of a gas-piston system, we cast them in a dimensionless form and we solve the dimensionless equations with the multiple scales expansion method. With the approximated  solutions we obtained, we express in a closed form the heat exchanged by the gas-piston system with the reservoir for a large class of relevant non-equilibrium situations.
\end{abstract}

\section{Introduction}

\begin{figure}
\centering
\includegraphics[width=0.6\linewidth]{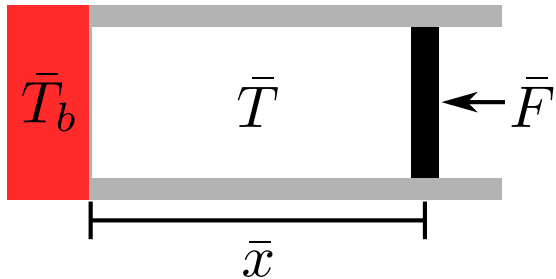}
\caption{A schematic representation of the gas 
canister with the macroscopic variables required to describe 
its dynamic.}\label{fig:gascanister}
\end{figure}

One of the irrefutable facts in learning thermodynamics
and statistical mechanics is that, when facing the 
two topics for the first time, one has to get skilled 
in solving problems that involve gasses and pistons. 
This is due to historical reasons. One is 
that thermodynamics was initially developed to understand  
the phenomenology of steam engines. One other is 
that the first model where the macroscopic properties 
of a body were linked to its internal constituents is 
the kinetic theory of gasses by Maxwell and Boltzmann. 
More than one hundred years have now passed form the 
work of Maxwell and Boltzmann and thermodynamics and 
statistical mechanics have widely changed: from theories 
that describe equilibrium conditions only, they are slowly
 but steadily evolving to include   non-equilibrium frameworks \cite{degrootmazur,balescu,sekimoto,seifert,campisi1} and results 
\cite{jarzynski,crooks,espositovandenbroeck,sano,campisi2}. However, gasses and
pistons have never stopped playing a prominent role 
in such evolution as their nonequilibrium behavior is 
not yet fully understood. As a proof,  Elliot Lieb 
stated that he would like to see solved the adiabatic piston problem
\cite{lieb}. It can be 
formulated as follow: we take an insulating canister 
with two gasses inside that are separated by a perfectly 
insulating moving piston. If the two gasses are 
initially at different pressures and temperatures, 
how is the equilibrium condition approached? It turns 
out that it's easy to answer this question qualitatively 
\cite{feynman}, while quantitative answers are still a 
very active field of research \cite{gruber,vulpiani3,cerino}. 
Curiously enough, the adiabatic piston problem is 
very similar to another which is not that much 
investigated, although it involves a conceptually 
simpler system and it's more relevant for applications. 
We consider the simplest thermodynamic machine: 
a perfect gas enclosed by a cylindrical canister 
with a movable piston and in contact with a heat 
reservoir (Figure \ref{fig:gascanister}). This system is simpler than the adiabatic piston as 1) only a single perfect gas is involved and 2) gas-reservoir heat exchanges are easily modeled microscopically \cite{tehver}.
Additionally, this 
device can be described by a limited set of 
macroscopic variables: the piston position 
$\overline{x}$, the gas internal temperature 
$\overline{T}$, the reservoir temperature 
$\overline{T}_{b}$ and the external force 
$\overline{F}$ applied on the piston. Among those 
variables, $\overline{F}$ and $\overline{T}_{b}$ 
can be changed according to some external time 
dependent protocol while $\overline{x}$ and 
$\overline{T}$ evolve as a consequence.
It is worth to note that, in the adiabatic piston problem, the force exerted by the piston can never be considered as an external known function, as it is an internal variable.
\\
One possible way to describe the time evolution of 
$\overline{x}$ and $\overline{T}$ is given in 
\cite{vulpiani1} through the following Gas-Piston equations\footnote{An anonymous referee pointed out that our eq.\eqref{eq:x_GPE} and the original eq. A1 in \cite{cerino} don't have the same last term. Such discrepancy is due to a confirmed misprint\cite{misprint} in \cite{cerino}. } (GPE)
\begin{subequations}
\begin{align}
    &
    \begin{aligned}
\ddot{\overline{x}}&+\frac{\overline{F}}{M}-\dfrac{\nu N}{\nu+1}\frac{1}{\overline{x}} \erfceqdim (\dot{\overline{x}}^2+\dfrac{\overline{T}}{m})\\
& + \dfrac{\nu N}{\nu+1}\expeqdim\frac{\dot{\overline{x}}}{\overline{x}}\sqrt{\dfrac{2\overline{T}}{\pi m}}=0,\\
    \end{aligned}\label{eq:x_GPE}
    \\
    &
    \begin{aligned}
        \dot{\overline{T}}&+\frac{2\dot{\overline{x}}\left[m\dot{\overline{x}}^2+\overline{T}(1-2\nu)\right]}{\overline{x}(\nu+1)^2} \erfceqdim+\sqrt{\dfrac{2\overline{T}}{\pi m}}\frac{\overline{T}-\overline{T}_b}{\overline{x}}\\
& +\frac{2}{\overline{x}(\nu+1)^2}\sqrt{\dfrac{2m\overline{T}}{\pi}}\left(\dfrac{2\nu\overline{T}}{m}-\dot{\overline{x}}^2\right)\expeqdim=0,
    \end{aligned}
\end{align}
    \label{eq:GPE}
\end{subequations}
where the upper dots denote time derivatives, $M$ 
is the mass of the piston, $m$ is the mass of a 
single gas molecule, $N$ is the total number of 
gas molecules, $\nu=m/M$ and $\erfc(\cdot)$ is the 
complementary error function.
Without going to much into details, these macroscopic equations are obtained by averaging microscopic properties with the aid of heavy assumptions. The first one is that the piston and each gas particle undergoes elastic collisions, so work is the energy exchanged in this way. The second assumption is that the velocity of a gas particle is randomly changed according to the Maxwell-Boltzmann distribution of the reservoir when reservoir-gas particle collisions occurs \cite{tehver}. Heat is no more than the change in energy of the gas due to this collision mechanism. The third assumption is that the gas distribution is always Maxwellian although gas-reservoir and gas-piston collisions change the gas temperature $\overline{T}$ over time. From a physical point of view, this third assumption rules out any shock-wave propagation, making the gas an efficient macroscopic dissipative medium \cite{vulpiani3}.
 
If the solution of 
eq.\eqref{eq:GPE} is known, it's possible to compute 
the total energy of the system $\overline{E}$ \cite{vulpiani1,jarzynski2}, 
the work $\overline{W}$ performed on the piston 
\cite{seifert,sekimoto}\footnote{The sign conventions 
for $\overline{Q}$ and $\overline{W}$ are the same used in \cite{seifert}, 
i.e. $\overline{Q}>0$ for heat given to the reservoir and $\overline{W}>0$ 
for work applied on the system.} and the heat $\overline{Q}$ 
exchanged with the reservoir as:
\begin{subequations}
	\begin{align}
		\overline{E}&=M\frac{\dot{\overline{x}}^2}{2}+\overline{F}\ \overline{x}+\frac{N									\overline{T}}{2},\\
		\overline{W}(\overline{t}_i,\overline{t}_f)&=\int_{\overline{t}_i}^{\overline{t}_f}									\dot{\overline{F}}\ \overline{x}\ \mathrm{d}\overline{t},\label{eq:jarzinskywork}\\
		\overline{Q}(\overline{t}_i,\overline{t}_f)&=-\int_{\overline{t}_i}^{\overline{t}_f} \bigg[  		\overline{F}\dot{\overline{x}}+ M\dot{\overline{x}}\ \ddot{\overline{x}} +\frac{N}{2}								\dot{\overline{T}}  \bigg]\, \mathrm{d}\overline{t}.
	\end{align}
    \label{eq:Energetics_dim}
\end{subequations}
It is worth to note that $\overline{F}\overline{x}$ is considered as a part of the total energy of the system. As a consequence, work must be defined as eq.\eqref{eq:jarzinskywork} \cite{jarzynski2}.

In this paper we present eq.\eqref{eq:GPE} in a new dimensionless form which allows to easily take the thermodynamic limits 
\begin{equation}
\nu\to0, \quad N\to\infty,
\label{eq:therm_lim}
\end{equation}
and to isolate 
\begin{equation}
\epsilon=\lim_{\substack{\nu\to0\\N\to\infty}}\sqrt{ \nu N},
\label{eq:epsdef}
\end{equation}
as the only relevant free parameter\cite{gruber}. Assuming that $\varepsilon$ is small and that the external force is slowly-varying over time, we linearize eq.\eqref{eq:GPE} and then proceed to an asymptotic expansion using the \emph{multiple scales method}. This way we find and approximated solution of the linearized equations which contains all the relevant physical behaviors of the system. Furthermore we use the obtained solutions to build closed form expressions for the heat exchanged with the reservoir for two relevant non-equilibrium transformations, namely the relaxation toward equilibrium and the isothermal compression of a gas realized in a finite time. In the next Section we give an account of the multiple scales method and then proceed with the outline presented above. 

\section{Short review on multiple scales method}
The history of the multiple scales method dates back to the 18th century from the works of
by Lindstedt \cite{Lindstedt1882} and Poincar\'e \cite{Poincare1886},
and was developed in its modern form in \cite{Kuzmak1959,ColeKevorkian1963}.
The core of the method is to find asymptotic approximated solutions to a differential equation 
when the standard perturbation theory produces secular terms.
During the years the multiple scales method has proven to be very useful in the
construction of approximate solutions of differential equations
and is now included in every textbook on perturbation theory 
\cite{Nayfeh,BenderOrszag,KevorkianCole,Holmes}.
Such powerful method has found applications also
in fields which seems not to have any correlation with such problems, for example
in the theory of integrable systems 
\cite{ZakharovKuznetsov1986,CalogeroEchkhaus1987,CalogeroEchkhaus1988,Calogero1991}. 

The key feature that allows the elimination of the secular terms
is the introduction of fast-scale variables and slow-scales variables
in a way that the dependence on the slow-scale variable will
prevent the secularities. To be more precise,
suppose that we are in the case
of an ordinary differential equation with respect to the independent
variable $t$ and a single dependent variable $x=x\left( t \right)$:
\begin{equation}
    E_{\varepsilon}\left( t,x,\dot{x},\ddot{x},\dots,\ndot{x}\right) =0,
    \label{eq:multiscalegen}
\end{equation}
where the $\varepsilon$ subscript means that we have
some dependence on a small parameter $\varepsilon$.
{We now assume that $x$ has an asymptotic series of the form:
\begin{equation}
    x\left( t \right)
    = \sum_{i=0}^{N+M} \varepsilon^{i} x_{i}\left( t_{0},t_{1},t_{2},\dots, t_{N} \right)+\BigO{\epsilon^{N+M+1}},
    \label{eq:asymptseries}
\end{equation}
truncated at some positive integer $N+M$, with $M\geq0$. In the right hand 
side of \eqref{eq:asymptseries} the dependence on the time variable $t$ 
appears trough the so  called \emph{scales} \footnote{If $t$ is a time 
variable, the scales are the characteristic 
\emph{time scales} of $x$. Similarly, if $t$ is 
a length variable, the scales are the characteristic 
\emph{length scales} of $x$} ${t_{i}=t_{i}(t,\varepsilon)}$. 
Intuitively, the scales isolate different behaviors inside eq. \eqref{eq:multiscalegen}, e.g. in the damped harmonic oscillator they separate oscillations from the amplitude suppression.
The number of scales to be introduced depends on the desired  asymptotic approximation 
order: the expansion is guaranteed to be asymptotic
until 
\begin{equation}
    t_{N}\left( t,\varepsilon \right) = \BigO{1}
    \label{eq:scalemeaning}
\end{equation}
is satisfied.
The number of scales also sets the approximation error,
in the sense that the maximum 
discrepancy from the complete solution
\begin{equation}
    \max_{t\in \left[ 0,t_{max} \right]}\abs{x(t)-
        \sum_{i=0}^{N+M-1} \varepsilon^{i} x_{i}\left( t_{0},t_{1},t_{2},\dots, t_{N} \right)}
    \label{eq:maxfunc}
\end{equation}
is $\BigO{\varepsilon^{N+M}}$, where $t_{max}$ 
is the time such that the condition \eqref{eq:scalemeaning}
holds.

The mathematical structure of the scales is the most delicate point 
in the whole expansion method: it involves the knowledge of 
eq.\eqref{eq:multiscalegen} structure, and the constraint that they 
must be \emph{non-decreasing} functions of $t$ which satisfy the 
condition:
\begin{equation}
    \lim_{\varepsilon\to0}\dfrac{t_{i+1}(t,\varepsilon)}{t_{i}(t,\varepsilon)} = 0, 
    \quad \forall i=0,1,\dots, N-1.
    \label{eq:timescalecond}
\end{equation}
{We note that $N$ has to be sufficiently high not just to give a longer asymptotic range of validity of the expansion, but also to capture the behavior of the system.}

The substitution \eqref{eq:asymptseries} can be extended
to all the derivatives of $x$ by differentiation, or more operatively
by substituting 
\begin{equation}
    \frac{\ud}{\ud t} \to \sum_{i=0}^{N}\pdv{t_{i}}{t}\pdv{}{t_{i}}.
    \label{eq:asymptder}
\end{equation}
Substituting eq.\eqref{eq:asymptseries} and all its 
derivatives in eq.\eqref{eq:multiscalegen}, eventually expanding in Taylor series with respect to $\epsilon$, we obtain a polynomial in $\varepsilon$
which is identically equal to zero. We can then separately set to zero all the coefficients of $\varepsilon$-powers and obtain a system of $N+M+1$ partial differential 
equations. If the scales are correctly  chosen, the  $\varepsilon^0$-equation,  will contain $x_{0}$ only and will depend just on $t_{0}$. This will
give raise to a solution depending on arbitrary functions of the remaining
scales $t_{1},\dots,t_{N}$. Substituting $x_{0}$ into the $\varepsilon^1$-equation  
 we  use these arbitrary functions to prevent the birth
of the secular terms in $x_{1}$. 
Solving iteratively for the remaining $x_{i}$ one finally writes
down the $N+M$ terms of the wanted expansion \eqref{eq:asymptseries}.
In the case of high order expansions ($N>1$) sometimes
the previous iterative method is not sufficient to completely specify the terms
of the asymptotic series. In these cases the strategy 
of the \emph{suppression of the order mixing} is adopted:
it consist in eliminating from the $\varepsilon^{i+1}$-equation all the contributions coming
from the arbitrary functions coming from lower orders solutions $x_{i}$, $x_{i-1}$, etc. This increases the accuracy of the first $i$ terms by reducing
the amounts of corrective terms in $x_{i+1}$ \cite{Holmes}.

%##################################################################################################
\section{Adimensionalization and expansion of the piston-gas equations}
\label{sec:dimnsionless_eqs}
Going back to the main aim of this paper, we introduce a dimensionless version of eq.\eqref{eq:GPE}, namely 
\begin{subequations}\label{eq:DGPE}
\begin{align}
    &\begin{aligned}
        \ddot{x}&+F+\erfceqadim\frac{\epsilon^2\dot{x}^2+T}{x}
        +\expeqadim\frac{\dot{x}}{x}\sqrt{\dfrac{2T}{\pi}}\epsilon=0,
    \end{aligned}
    \\
    &\begin{aligned}
        \dot{T}&-2\erfceqadim \frac{\dot{x}}{x}(\epsilon^2\dot{x}^2+T)+\sqrt{\dfrac{2T}{\pi}}\frac{T-T_b}{\epsilon x}
        \\
        &-2\sqrt{\dfrac{2T}{\pi}}\epsilon\frac{\dot{x}^2}{x}\expeqadim=0.
    \end{aligned}
\end{align}
\end{subequations}
To obtain this expression we have introduced in eq.\eqref{eq:GPE} the dimensionless quantities 
$x$, $T$, $F$, $T_b$ and the dimensionless time $t$ via
\begin{subequations}
\begin{align}
\overline{x}&=x\cdot \frac{T_{br}N}{F_r(1+\nu)g(\nu)}\\
\overline{T}&=T\cdot \frac{T_{br}}{g(\nu)}\\
\overline{F}&=F\cdot F_{r}\\
\overline{T}_b&=T_b\cdot T_{br}\\
\overline{t}&=\frac{t}{F_{r}}\cdot \sqrt{\frac{mNT_{br}}{\nu(1+\nu)g(\nu)}} 
\end{align}
    \label{eq:dimensionless_variables}
\end{subequations}
where
\begin{equation}
g(\nu)=\frac{1+6\nu+\nu^2}{(1+\nu)^2}
\end{equation}
and $F_r$ ($T_{br}$) is an arbitrary force (temperature) reference value. Similarly we write the dimensionless $E$, $W$ and $Q$ densities:
\begin{subequations}
\begin{align}
E=&\frac{\overline{E}}{NT_{br}}=Fx+\frac{\dot{x}^2}{2}+\frac{T}{2},\\
W(t_i,t_f)=&\frac{\overline{W}}{NT_{br}}=\int_{t_i}^{t_f}\dot{F}x \ \mathrm{d}{t},\\
Q(\overline{t}_i,\overline{t}_f)=&\frac{\overline{Q}}{NT_{br}}=-\int_{t_i}^{t_f} \bigg[  F\dot{x}+ \dot{x}\ddot{x} +\frac{1}{2}\dot{T}  \bigg]\, \mathrm{d}t.\label{eq:Energetics_adim_Q}
\end{align}
    \label{eq:Energetics_adim}
\end{subequations}
It is important to note that in eq.(\ref{eq:DGPE}) and (\ref{eq:Energetics_adim}) the thermodynamic limit eq.\eqref{eq:therm_lim} are already taken, provided that the gas-piston mass ration $\varepsilon$ defined in eq.\eqref{eq:epsdef} is finite. We also note that 
\eqref{eq:DGPE} and \eqref{eq:Energetics_adim} 
show an important property firstly observed 
in the related adiabatic piston problem \cite{gruber}: 
the sole knowledge of $\epsilon$ is sufficient 
to describe the general features of
the system while the remaining parameters 
appears as scaling factors. Following again \cite{gruber}
we also treat $\varepsilon$ as a small perturbation parameter.

We now restrict ourselves to the case where the 
reservoir temperature is constant over time 
while $F$ evolves according to a given slow protocol. 
W.l.o.g, we take ${T_b=1}$. $F$ can be considered slow if 
\begin{equation}
\pdv{F}{t}\propto \epsilon^g \quad \text{for } g\geq1. \label{eq:slownesscondition}
\end{equation}
The simplest way to satisfy eq.\eqref{eq:slownesscondition} is to take $F=F(\epsilon t)$, which we assume from now on.

Since $F$ varies slowly, the equilibrium point of eq.\eqref{eq:DGPE}, namely
\begin{equation}
    x_{\text{eq}}=\frac{1}{F}, \quad \dot{x}_{\text{eq}}=0, \quad T_{\text{eq}}=1,
    \label{eq:equil}
\end{equation}
is also slowly varying over time. Since we are interested in results of thermodynamic relevance, we are allowed to linearize eq.\eqref{eq:DGPE} around eq.\eqref{eq:equil}, which yields
\begin{subequations}
\begin{align}
    &\ddot{x}+{F^2(\epsilon t)}\, x +2\sqrt{\dfrac{2}{\pi}}\epsilon F(\epsilon t)\, \dot{x}-F(\epsilon t)\, T=0,
    \label{eq:sysxT1_eqx}
\\
&\dot{T}+2F(\epsilon t)\,\dot{x}+\sqrt{\dfrac{2}{\pi}}\frac{F(\epsilon t)}{\epsilon}\,(T-1)=0.
\label{eq:sysxT1_eqT}
\end{align}\label{eq:sysxT1}
\end{subequations}
We call eq.\eqref{eq:sysxT1} the Linearized Dimensionless Gas Piston Equations (LDGPE).

Being the LDPGE linear,  we could
expect an exact analytic solution. However Computer 
Algebra Systems like \texttt{Macsima} or \texttt{Maple} 
show that analytic solutions of the LDGPE are 
of no practical use or even impossible to express
 because they are too strongly 
dependent on $\varepsilon$ value and on the 
functional form of $F$. 

The LDGPE equations \eqref{eq:sysxT1} are a system of a second
order equation coupled with a first order one. However we show that it can be treated as a single third order equation. We solve
eq. \eqref{eq:sysxT1_eqx} for $T$:
\begin{equation}
    T=\frac{\ddot{x}}{F(\varepsilon t)}+F(\varepsilon t)x+2\sqrt{\tfrac{2}{\pi}}\epsilon\dot{x}
    \label{eq:solution_T_0}
\end{equation}
and then insert it into \eqref{eq:sysxT1_eqT} to obtain:
\begin{equation}
    \begin{aligned}
        \frac{\dddot{x}}{F \left( \epsilon t \right) }&+
        \left[ {\frac {2\sqrt {2}\epsilon}{\sqrt {\pi }}}
        +{\frac {\sqrt {2}}{\sqrt {\pi }\epsilon}}
        -{\frac {F' \left( \epsilon t \right) \epsilon}{F^{2} \left( \epsilon t \right)}} \right]
        \ddot{x}         
        + \left[ 3 F \left( \epsilon t \right) +{\frac {4F \left( \epsilon t \right) }{\pi }}\right]
        \dot{x}
        \\
        &+ \left[ F' \left( \epsilon t \right) \epsilon
        +{\frac {\sqrt {2} F^{2} \left(\epsilon t \right)}{\sqrt {\pi }\epsilon}} \right] x
        -{\frac {\sqrt {2}F \left( \epsilon t \right) }{\sqrt {\pi }\epsilon}}=0,
    \end{aligned}
    \label{eq:x_iii}
\end{equation}
where $'$  stands for the differentiation with respect to the
argument\footnote{In general $G'(j)=\left.\dv{G(\mu)}{\mu}\right|_{\mu=j}$ where $\mu$ is a dummy variable.}. 

Since the force is slowly varying and we don't want to treat
just a particular case, we are naturally led to consider
not a standard perturbative approach\footnote{Expanding in
the standard way requires to express $F(\varepsilon t)$ as Taylor
series, thus losing generality.}, but the multiple scales one.

The $\varepsilon$-perturbation is singular, i.e.
if $\varepsilon\to0$ then \eqref{eq:x_iii} ceases to be a third
order equation, but collapse into a second order one. Indeed
from \eqref{eq:x_iii} it is clear that as $\varepsilon\to0$
the dominant term is given by:
\begin{equation}
    \ddot{x} + F^{2}\left( \varepsilon t \right) x = F\left( \varepsilon t \right)
    \label{eq:x_iii_sing}
\end{equation}
which is just an harmonic oscillator with slowly varying frequency
$\omega(t) = F\left( \varepsilon t \right)$ and slowly forcing 
$\mathcal{F}(t)=F\left( \varepsilon t \right)$. To avoid the singularity we just make an $\varepsilon$-scaling in $x$ and $t$,
with undetermined coefficients, i.e. $x\left( t \right) = 
\varepsilon^{\alpha} X\left( \varepsilon^{\beta}t \right)$. Substituting
into \eqref{eq:x_iii} we found that we have to impose $\alpha=0$ and
 $\beta=-1$; putting $\tau=t/\varepsilon$ we obtain:
\begin{equation}
    \begin{aligned}
        &\phantom{+}\frac {1}{F \left( {\epsilon}^{2}\tau \right) }\dv[3]{X}{\tau}
        +\sqrt{\frac {2}{\pi}}\dv[2]{X}{\tau}
        \\
        &+{\epsilon}^{2}
        \left\{\left[2 \sqrt{\frac {2}{\pi}} 
        -\frac { F'\left( {\epsilon}^{2}\tau \right) }{F^{2} \left( {\epsilon}^{2}\tau \right)}\right]
        \dv[2]{X}{\tau}
        +\left( 3 + \frac{4}{\pi} \right)  F \left( {\epsilon}^{2}\tau \right)\dv{X}{\tau}\right.
        \\
        &\left.
        +\sqrt {\frac {2}{\pi}}F^{2} \left( {\epsilon}^{2}\tau \right)X
        -\sqrt {\frac{2}{\pi}} F \left( {\epsilon}^{2}\tau \right)\right\} 
        +\epsilon^{4} F'\left( {\epsilon}^{2}\tau \right) X=0.
    \end{aligned}
    \label{eq:X_iii}
\end{equation}
As can be easily seen now we have no singularity as $\varepsilon\to0$
and the dominant term is now a third order equation.

The next step is to introduce the time scales.
 It is easy to see that if we choose the 
trivial time scales 
\begin{equation}
    \btau=(\tau_{0},\tau_{1},\tau_{2},\dots)=
    (\tau,\varepsilon\tau,\varepsilon^{2}\tau,\dots)
    \label{eq:trival}
\end{equation}
we end up with an asymptotic expansion which is identically zero, 
meaning that this choice is not correct. However,
we can use the trivial time scales when the force $F\left( \varepsilon^{2}\tau \right)\equiv1$. 
To construct the time scales in 
the general setting $F\left( \varepsilon^{2} \tau \right)\not\equiv1$, we search 
 a change of variables $\tau\to\tau_{0}\left( \tau\right)$
such that the $\epsilon^0$ term in eq.\eqref{eq:X_iii} reduces to the
$\epsilon^0$ term for the $F\left( \varepsilon^{2}\tau \right)\equiv1$ case.
Performing such change of variables we obtain:
\begin{equation}
    \begin{aligned}
        &\phantom{+}X'''\left( { \tau_{0}} \left( \tau \right)  \right)  \left(\tau_{0}' \left( \tau \right)  \right) ^{3}
        +3 X''  \left( { \tau_{0}} \left( \tau \right)  \right) { \tau_{0}'} \left( \tau \right)
         \tau_{0}''\left( \tau \right) 
        +X'  \left( \tau_{0}\left( \tau \right)  \right) \tau_{0}'''\left( \tau \right)
        \\ 
        &+ \sqrt {\frac{2}{\pi}} F \left( {\epsilon}^{2}\tau \right)
        \biggl[  X''  \left( { \tau_{0}} \left( \tau \right)  \right)  \left(\tau_{0}' \left( \tau \right)  \right) ^{2}
       	+X'  \left( { \tau_{0}} \left( \tau \right)  \right) \tau_{0}'' \left( \tau \right)  \biggr] 
        + \BigO{\varepsilon^{2}}=0.
    \end{aligned}
    \label{eq:calctimescale}
\end{equation}
Since the $\varepsilon^0$ term for  $F\left( \varepsilon^{2}\tau \right)\equiv1$ is 
\begin{equation}
    \epsilon^0\colon \dv[3]{X}{\tau} + \sqrt{\frac{2}{\pi}}\dv[2]{X}{\tau}
    \label{eq:lot1}
\end{equation}
we see that we have:
\begin{equation}
    \tau_{0} = \int_{0}^{\tau} F\left( \varepsilon^{2} \chi \right) \ud \chi.
    \label{eq:tau0}
\end{equation}
For the second scale we repeat the same procedure as $\tau_{1}=\varepsilon\tau$
implies an identically zero asymptotic behavior. 
Performing again such procedure we see that  the
second scale is $\tau_{1}=\varepsilon \tau_{0}$. As third scale we can
safely choose $\tau_{2} = \varepsilon^{2}\tau$.
In the end we have the following three scales:
\begin{equation}
    \btau = \left(\tau_0,\tau_1,\tau_2 \right)=
    \left(\int_{0}^{\tau} F\left( \varepsilon^{2} \chi \right) \ud \chi,
    \varepsilon\int_{0}^{\tau} F\left( \varepsilon^{2} \chi \right) \ud \chi,
    \varepsilon^{2}\tau
    \right).
    \label{eq:scales}
\end{equation}
It's worth to note that, two different time scales faster that the external driving time scale $\epsilon t$ are required for a full description of the system. Since the time scales of the multiple scales method are each one associated  to a different physical phenomena, eq.\eqref{eq:scales} gives us a rigorous proof that $F=F(\epsilon t)$ is indeed a slow force if compared with the remaining fundamental time scales of the system. We will address to which phenomena $\tau_0$ and $\tau_1$ are related later in this paper. We note that the condition \eqref{eq:timescalecond}
is satisfied by the scales \eqref{eq:scales},
but the requirement for them to be non-decreasing functions
impose some restrictions on $F$.
We note that if function $F$
is always \emph{positive}, then this requirement
is automatically satisfied. Many cases of physical
interests satisfies such positivity requirement and these
will be discussed later in the paper.

Now we introduce the truncated expansion:
\begin{equation}
    X(t) = X_{0}(\btau) + \varepsilon X_{1}(\btau)
    +\varepsilon^{2}X_{2}(\btau) + \BigO{\varepsilon^{3}},
    \label{eq:Xap}
\end{equation}
and substitute it into \eqref{eq:X_iii} with $\btau = \left(\tau_0,\tau_1,\tau_2 \right)$. Taking the coefficients with respect to $\varepsilon$ to be zero,
we found the following equations up to $\varepsilon^{2}$:
\begin{subequations}
    \begin{align}
        \varepsilon^{0} &\colon
        \phantom{+}\pdv{X_{0}}{*{3}{\tau_{0}}}+\sqrt{\frac {2}{\pi}}\pdv{X_{0}}{*{2}{\tau_{0}}}=0,
        \label{eq:c0}
        \\
        \varepsilon^{1} &\colon
        %\begin{aligned}[t]
           \phantom{+}\frac {\partial ^{3} X_{1}}{\partial {\tau_{0}}^{3}}
        +\sqrt \frac{2}{\pi}\frac {\partial ^{2} X_{1}}{\partial {\tau_{0}}^{2}}
        %\\
        +2 \sqrt {\frac{2}{\pi }}{\frac {\partial ^{2} X_{0}}{\partial {\tau_{1}}\partial {\tau_{0}}}}
        +3 \frac {\partial ^{3} X_{0}}{\partial {\tau_{1}}\partial {{\tau_{0}}}^{2}}=0,
        %\end{aligned}
        \label{eq:c1}
        \\
        \varepsilon^{2} &\colon
        \begin{aligned}[t]
            &\phantom{+}\frac {\partial ^{3} X_{2}}{\partial \tau_{0}^{3}}
            +\sqrt \frac{2}{\pi}\frac {\partial ^{2} X_{2}}{\partial \tau_{0}^{2}}
            +3 \frac {\partial^{3} X_{1}}{\partial \tau_{1}\partial \tau_{0}^{2}}
            +2\sqrt{\frac{2}{\pi}} {\frac {\partial ^{2} X_{1}}{\partial {  \tau_{1}}\partial {  \tau_{0}}}}
            \\
            &+\sqrt{\frac{2}{\pi}}\frac {\partial ^{2} X_{0}}{\partial \tau_{1}^{2}}
            +3 \frac{\partial ^{3} X_{0}}{\partial \tau_{1}^{2}\partial \tau_{0}}
            +\left(3+\frac {4}{\pi }  \right) \frac {\partial X_{0}}{\partial \tau_{0}}
            +2 \sqrt{\frac{2}{\pi}}\frac {\partial ^{2} X_{0}}{\partial \tau_{0}^{2}}
            \\
            &+\sqrt{\frac{2}{\pi}} X_{0}
            +\frac {3}{F \left(\tau_{2}\right)}\frac{\partial ^{3} X_{0}}{\partial \tau_{2}\partial \tau_{0}^{2}}
            +\sqrt{\frac{2}{\pi }}\frac {2}{F \left( {\tau_{2}} \right)}
            {\frac {\partial ^{2} X_{0}}{\partial {  \tau_{2}}\partial {  \tau_{0}}}}
            -\sqrt{\frac{2}{\pi}}\frac {1}{F \left(\tau_{2} \right)}
            \\
            &+2 \frac { F' \left( {  \tau_{2}} \right) }{ F^{2}\left( \tau_{2} \right)}
            \frac{\partial ^{2} X_{0}}{\partial \tau_{0}^{2}}
            +\sqrt{\frac{2}{\pi}}\frac { F' \left( {  \tau_{2}} \right) }{ F^{2}\left( \tau_{2} \right)}
            {\frac {\partial X_{0}}{\partial {  \tau_{0}}}}=0.
        \end{aligned}
    \end{align}
    \label{eq:coeffseps}
\end{subequations}
It is very easy to see that the solutions to these equations 
are \emph{weakly secular} in the sense that, except in some
notable cases, the convergence of expansion is ensured by the presence
of exponentially decreasing functions. Therefore we adopt the strategy of the suppression of order mixing: we use the arbitrary functions in $X_0$, \dots, $X_i$ to eliminate as much as possible the presence of  $X_0$, \dots, $X_i$ in the equations for $X_{i+1}$, \dots, $X_{N+M}$.
At this point we notice that to give a 
\emph{complete characterization} to the function $X_{0}$, $X_{1}$ and $X_{2}$ 
it is not possible to just use the three equations above, but one must add terms
up to $\varepsilon^6$. We omit the further three equations and all the calculations, since they are very long,
but in fact trivial.
The results of the calculations, once written in the original variable
time scales
\begin{equation}
    \mathbf{t} =
    \left(\frac{1}{\varepsilon}\int_{0}^{t} F\left( \varepsilon \chi \right) \ud \chi,
    \int_{0}^{t} F\left( \varepsilon \chi \right) \ud \chi,
    \varepsilon t\right)
    \label{eq:scalest}
\end{equation}
becomes:
\begin{equation}
    x(t) 
    \begin{aligned}[t]
        &=
        \frac{1}{F\left( \epsilont \right)}
        +\frac{K_{1}}{F\left( \varepsilon t \right)}
        e^{\frac{1}{\varepsilon}\left( \sqrt{2\pi}-\sqrt{\frac{2}{\pi}} \right)\IF}
        \\
        &+\frac{e^{-\varepsilon \left(\sqrt{\frac{2}{\pi}}+\sqrt{\frac{\pi}{2}}  \right)\IF}}{%
            \sqrt{F\left( \varepsilont \right)}}\left[C_{1}\sin\left( \IF \right) + C_{2} \cos\left( \IF \right) \right]
        \\
        &
        +\varepsilon\left\{ 
            \frac{e^{-\varepsilon \left(\sqrt{\frac{2}{\pi}}+\sqrt{\frac{\pi}{2}}\right)\IF}}{%
                \sqrt{F\left( \varepsilont \right)}}
                \left[\left( C_{3}+ C_{2}\Theta\left( \varepsilon t \right) \right)
                \sin\left( \IF \right)
                \right.
            \right.
                \\
                &\left.
                \left.+\left( C_{4} - C_{1}\Theta\left( \varepsilon t \right) \right)
                \cos\left( \IF \right)
            \right]
         +\frac{K_{2}}{F\left( \varepsilon t \right)}
        e^{\frac{1}{\varepsilon}\left( \sqrt{2\pi}-\sqrt{\frac{2}{\pi}} \right)\IF}
        \right\}
        \\
        &
        +\varepsilon^{2}x_{2}\left( \mathbf{t};\left\{C_{i},K_{3}\right\} \right) 
        + \BigO{\varepsilon^{3}}.
    \end{aligned}
    \label{eq:solution}
\end{equation}
Where:
\begin{equation}
    \begin{aligned}
        \Theta(s) &= \int_{0}^{s}\left[ 
            \left( \frac{1}{\pi }+1-\frac{\pi}{4}  \right) F \left( s \right) 
            \right.
            %\\
            %&
            + \frac{1}{2}\left(\sqrt{\frac{2}{\pi}}+\sqrt{\frac{\pi}{2}} \right)  
            \frac{F'(s)}{F \left( { s} \right)}
            \\
            &\phantom{= \int_{0}^{s}}
            \left.+\frac{1}{4}\frac {F''\left( { s} \right) }{F^{2}\left( s \right)}
            -\frac{3}{8}\frac {  (F')^{2}  \left( { s} \right)}{ F^{3} \left( { s} \right)}
        \right]
        \ud s
    \end{aligned}
    \label{eq:wfunc}
\end{equation}
The $x_{2}$ part of the function is not displayed in its generality
since it is very long and cumbersome, but we note that the writing 
$x_{2}=x_{2}\left( \mathbf{t};\left\{C_{i},K_{j}\right\} \right)$
means that $x_{2}$ has parametric dependence on the
$C_{i}$, $i=1,2,3,4,5,6$ and $K_{3}$ which are the
constants of integration.
In the next sections, while discussing some
particular cases of $F$, we show the specific forms it assumes.

We can construct a general formula for the
expression of the asymptotic series for $T$ substituting eq.\eqref{eq:solution}
into eq.\eqref{eq:solution_T_0}:
\begin{equation}
    T(t) 
    \begin{aligned}[t]
        &= \frac{F(\epsilon t)}{\varepsilon^{2}} \pdv{x_{0}}{*{2}{t_{0}}}
        +\frac{F\left( \varepsilon t \right)}{\varepsilon}\left( \pdv{x_{1}}{*{2}{t_{0}}}
        + 2 \pdv{x_{0}}{ {t_{0}}{t_{1}}}\right)
        %\\
        %&
        +F \left( \varepsilon t \right) x_{0} 
        \\
        &
        +2\sqrt{\frac{2}{\pi }}
        F \left( \varepsilon t \right)\frac {\partial x_{0}}{\partial {t_0}}
        +2{\frac {\partial ^{2} x_{0}}{\partial { t_{0}}\partial { t_{2}}}}
        +F \left( \varepsilon t \right) {\frac {\partial^{2} x_{2}}{\partial { t_0}^{2}}}
        \\
        &
        +2 F \left( \varepsilon t \right) {\frac {\partial ^{2} x_{1}}{\partial { t_0} \partial { t_1}}}
        %\\
        %&
        +F \left(\varepsilon t \right)  \frac {\partial ^{2} x_{0}}{\partial {t_1^{2}}}
        +\frac{F' \left( \varepsilon t\right) }{F \left( \varepsilon t\right) }
        \frac {\partial x_{0}}{\partial { t_0}}
        +\BigO{\varepsilon}
    \end{aligned}
    \label{eq:solution_T}
\end{equation}
We have then that  the error on $T$ is of order $\varepsilon$ since
 to have a better estimate on it, it is
necessary the knowledge of the $x_{3}$  term.

A particularly interesting case arise when all the constants of
integration are taken to be zero, $C_{i}=K_{j}=0$, which, being
the system linear, corresponds to the case when the initial condition
is trivial and the system evolves according to the external forcing.
Since the system is, in general, non-autonomous this is the
\emph{dynamical equilibrium} and is given by:
\begin{equation}
    x_{\deq}(t)
    %\begin{aligned}[t]
        %&
        =\frac{1}{F\left( \varepsilon t \right)}
        + \varepsilon^{2} \left[ \frac{F''\left(\varepsilon t\right)}{F^{4}\left( \varepsilon t \right)}
            -2 \frac{(F')^{2}\left(\varepsilon t\right)}{F^{5}\left( \varepsilon t \right)}
            %\right.
            %\\
            %&\left.
            + \left( 2\sqrt{\frac{2}{\pi}} + \sqrt{2\pi} \right)
            \frac{F'\left( \varepsilon t \right)}{F^{3}\left( \varepsilon t \right)} \right]
            +\BigO{\varepsilon^{4}}
    %\end{aligned}
    \label{eq:dynequil}
\end{equation}
Notice that as was previously known \cite{vulpiani1} 
the $\varepsilon^0$ contribution to the dynamical equilibrium solution is the same as that
of the underlying forced harmonic oscillator \eqref{eq:x_iii_sing}. The first
correction is then second order, while the next one will be at fourth\footnote{We remark
that to compute the complete expansion we needed six terms.}.

Upon differentiation with respect to $t$ from \eqref{eq:dynequil} we obtain 
the modified equilibrium conditions for $\dot x$ and $T$
(the latter by using \eqref{eq:solution_T}). For the
dynamical equilibrium $T$ we obtain from \eqref{eq:solution_T}
the following very simple expression:
\begin{equation}
    T_{\deq}(t) = 1 + \epsilon^2\sqrt{2\pi}\frac{F'(\epsilon t)}{F^2(\epsilon t)} +\BigO{\varepsilon^4}.
    \label{eq:dynequilT}
\end{equation}
We remark that upon substituting $F\equiv1$ the dynamical
equilibrium reduces exactly to the usual equilibrium condition
$x=1$, $\dot x =0$ and $T=1$. We also note that eq.\eqref{eq:dynequil} and \eqref{eq:dynequilT} can be derived as standard perturbation expansion assuming $x(t)=x_{0}(\epsilon t)+\epsilon x_{1}(\epsilon t)+ \epsilon^2 x_{2}(\epsilon t)+\epsilon^3 x_{3}(\epsilon t)+\BigO{\epsilon^4}$.

Since our starting hypothesis was that the original 
DGPE are in linear regime, it is particularly useful 
to express the initial conditions
of the system not as generic, but as deviation from the dynamical equilibrium
\eqref{eq:dynequil}. Using \eqref{eq:dynequilT} and \eqref{eq:solution_T_0}
valued at $t=0$ we find the following values for the near equilibrium 
initial conditions for eq. \eqref{eq:x_iii}:
\begin{subequations}
    \begin{align}
        x(0)
        &%\begin{aligned}[t]
            %&
            =x_{0} + \frac{1}{F\left( 0 \right)}
            + \varepsilon^{2} \left[ \frac{F'\left(0\right)}{F^{4}\left( 0 \right)}
                -2 \frac{(F')^{2}\left( 0\right)}{F^{5}\left( 0 \right)}
                %\right.
                %\\
                %&\left.
                + \left( 2\sqrt{\frac{2}{\pi}} + \sqrt{2\pi} \right)
                \frac{F'\left( 0 \right)}{F^{3}\left( 0 \right)} \right]
                +\BigO{\varepsilon^{4}}
        %\end{aligned}
        \\
        \dot x(0) &= \dot x_{0} - \varepsilon \frac{F'(0)}{F^{2}(0)}
        +\BigO{\varepsilon^{3}}
        \\
        \ddot{x}(0) &= 1 + T_{0}-x_{0}-\frac{1}{F(0)}  + \BigO{\varepsilon}.
    \end{align}
    \label{eq:ics}
\end{subequations}
Here $x_{0}$, $\dot x_{0}$ and $T_{0}$ are taken to be $\BigO{1}$ deviations from equilibrium eq. \eqref{eq:dynequil}.
This will give us the following values for the constants of integration:
\begin{equation}
    \begin{aligned}
        C_{1} &= \frac{\dot x_{0}}{\sqrt{F\left( 0 \right)}} 
		\\
        C_{2} &= x_{0} \sqrt{F\left( 0 \right)},
        \\        
        C_{3} &=
            \begin{gathered}[t]
             \left(\sqrt{2\pi} 
            + \sqrt{\frac{2}{\pi}}\right) \sqrt{F\left( 0 \right)} x_{0} 
            + \frac{1}{2} \frac{F'\left( 0 \right)}{F^{3/2}\left( 0 \right)}x_{0}
            \\
            + \frac{1}{2}\frac{\sqrt{2\pi}}{F^{3/2}\left( 0 \right)}\left( 1 + T_{0} -x_{0} \right)
            - \frac{1}{2} \frac{\sqrt{2\pi}}{F^{5/2}\left( 0 \right)}
        \end{gathered}
        \\
%        \begin{gathered}
            C_{5} &= \half\frac {\pi { x_{0}}}{ F^{3/2} \left( 0 \right)}
            -\half\frac {\pi }{ F^{3/2} \left( 0 \right)}
%            \\
            -\half\pi \sqrt {F \left( 0 \right) }{ x_{0}}
            +\half\frac {\pi }{ F^{5/2} \left( 0 \right)}
            -\half\frac {\pi { T_{0}}}{ F^{3/2} \left( 0 \right)} 
%        \end{gathered}
        \\
%        \begin{gathered}
            K_3 &= \frac {\pi }{2F \left( 0 \right) }
            +\frac {{ T_0}\pi }{2 F \left( 0 \right) }
            -\frac {\pi }{ 2 F^{2} \left( 0 \right)}
%            \\
            +\frac{1}{2}{x_0}\pi F \left( 0 \right) 
            -\frac {\pi x_0 }{2F \left( 0 \right) }
%        \end{gathered}
    \end{aligned}
    \label{eq:csfix}
\end{equation}
whereas $K_{1}=K_{2}=C_{4}=0$ and $C_{6}$ is left unspecified, meaning that
it can be safely put to zero.
We note that the initial conditions are satisfied exactly at $x(0)$,
up to order $\BigO{\varepsilon^{2}}$ at $\dot x(0)$ and up to order $\BigO{\varepsilon}$
at $\ddot x(0)$. This is not an accident of our system, but is
 a standard feature of the multiple scales approach
\cite{KevorkianCole}. The fact that $K_{1}=K_{2}=0$ is not surprising. From
eq. \eqref{eq:solution_T}
it is quite clear that the first two orders $\varepsilon^{-2}$ and $\varepsilon^{-1}$
must vanish to get as  initial condition $1+T_{0}=\BigO{1}$.

Without the need for an explicit form of $F$ 
we can now give an intuitive meaning to the the 
three scales we introduced. 
The $t_{0}$-scale is the fastest one and
characterizes an exponential-relaxation of the system
toward the equilibrium position given by \eqref{eq:dynequil}. 
We notice from $K_{1}=K_{2}=0$ that 
this scale appears in $x$  as an $\epsilon^2$ term,
 giving very little contribution.
We also note that, due to the presence of second order derivatives in eq.\eqref{eq:solution_T}
the $t_0$ scale appears in $T$ as the leading order term. This means that the temperature of the gas can have deviations from equilibrium as large as $\epsilon^{-2}$ and still rapidly converge to the equilibrium value. 
The $t_{1}$-scale is the one at which the oscillations of
the system are established. It is worth to note, that even if the system possess
a clear dissipative behavior, the basic frequency of the system
is unaltered adding the third scale meaning that, if any correction
in the basic frequency exists, then it should be at least of order 
$\varepsilon^{2}t$.
In the $t_{2}$-scale the exponential suppression
of the oscillatory terms appears; this means that the oscillations are slowly modulated.
Overall we see, under suitable assumption on the smallness of $\varepsilon$, that the approximated solution tends to the dynamical equilibrium
solution \eqref{eq:dynequil} as $t\to\infty$
which is coherent with a transient like behavior.
We note that using only two time scales would have resulted in missing the modulation
of the oscillation, leading to an erroneous result from both the physical and mathematical point of view. 
{ The previous considerations give us an \emph{a posteriori} justification
of the choice of using only three time scales, since all
the above features describe well the behavior of the system from both a numerical and a physical point of view}.

As a final recall on terminology we call from now  on $x_{\app}$ the truncated part of
the expansion for $x$ at order $\varepsilon^{3}$ given by \eqref{eq:solution},
and we call  $T_{\app}$ the truncated part of the expansion for $T$ 
at order $\varepsilon$. The next two Sections are devoted to two particular examples of thermodynamic relevance which we will use to test the quality of $x_{\app}$ as an approximated solution of the LDGPE and to derive new closed form expression of $Q$.

\section{Relaxation to equilibrium}
The first case under study is the one in which
\begin{equation}
	F=1\quad \forall t\geq0
	\label{eq:autonomous_force}
\end{equation}
with initial conditions
\begin{equation}
	\begin{aligned}
		x(0)=&1+x_0\\
		\dot{x}(0)=&\dot{x}_0\\
		T(0)=&1+T_0.
	\end{aligned}
	\label{eq:autonomous_BC}
\end{equation}
This simple case encompasses all the situations 
where the gas and the piston relax from a given 
nonequilibrium condition $\{1+x_0,\dot{x}_0,1+T_0\}$ 
to the thermodynamic equilibrium $\{x_{eq}=1$, 
$\dot{x}_{eq}=0$, $T_{eq}=1$\}. Substituting 
eq.\eqref{eq:autonomous_force} in eq.\eqref{eq:solution} 
and eq.\eqref{eq:scalest}, 
and then imposing that eq.\eqref{eq:csfix} must 
hold, we obtain an approximated expression for the piston position  
\begin{equation}
	\begin{aligned}
		x_{ap}(t)=&1+\exp\left( -\frac{\epsilon t (\pi+2)}{\sqrt{2\pi}} \right) \bigg[ C_1\sin(t)+C_2\cos(t)\\
			   &+\Big(-\frac{t\epsilon^3}{2}(\epsilon t \theta^2-2\eta)C_1+\theta t C_2									\epsilon^2+C_3 \epsilon\Big)\sin(t)\\
			   &+\Big(-\frac{\theta^2 t^2}{2} C_2 \epsilon^4+(C_2 \eta t-C_3 t \theta) \epsilon^3 +(C_5-C_1 t \theta) \epsilon^2\Big)\cos(t)\bigg]\\
			   &+K_3\epsilon^2 \exp\left(\sqrt{\tfrac{2}{\pi}}\frac{t}{\epsilon}  (\pi\epsilon^2-1) 			   \right)
	\end{aligned}
	\label{eq:autonomous_solution}
\end{equation}
with 
\begin{equation}
	\begin{aligned}
		\eta=&-\frac{\sqrt{2\pi}}{4} (\pi+4)\\
		\theta=&-\frac{(\pi^2-4 \pi-4)}{4 \pi}
	\end{aligned}
\end{equation}
and nonzero integration constants
\begin{equation}
	\begin{aligned}
		C_1=&\dot{x}_0\\
		C_2=&x_0\\
		C_3=&\frac{(\pi T_0+\pi x_0+2x_0)}{\sqrt{2\pi}}\\
		C_5=&-\frac{\pi T_0}{2}\\
		K_3=&\frac{\pi T_0}{2}.
	\end{aligned}
\end{equation}
In addition to the general properties of the scales inherited from eq.\eqref{eq:solution}, eq.\eqref{eq:autonomous_solution} shows two interesting properties. The first one is given by its general structure: being eq.\eqref{eq:autonomous_solution} made of a constant plus two decaying functions, it thermalizes and it describes well the relaxation to equilibrium. The second property is that we are now able to address the physical phenomena to which $t_0$ ($\tau_0$) and $t_1$ ($\tau_1$) are related to. As a matter of fact, $\tau_1$ gives the suppression mechanism related to the mechanical damping the gas acts on the piston with a characteristic time of $\frac{\sqrt{2\pi}}{\epsilon	(\pi+2)}$. On the other hand, $\tau_0$ gives a second suppression mechanism emerging from the indirect coupling of the piston with the reservoir with a characteristic time of $\frac{\epsilon}{\pi\epsilon^2-1}\sqrt{\frac{\pi}{2}}$. The fact that the piston-reservoir interaction is indirect is shown by the fact that this effect is $\BigO{\epsilon^2}$ in $x$ whereas in $T$ (where the gas-reservoir contact is direct) this effect is $\BigO{\epsilon^0}$. This feature is not surprising, as the the temperature reservoir appears explicitly only in eq.\eqref{eq:sysxT1_eqT} and not in eq.\eqref{eq:sysxT1_eqx}, but we are now able to describe quantitatively this phenomena.

To test the quality of $x_{ap}$ as solution we note 
that the exponents of eq.\eqref{eq:solution} converge only if 
\begin{equation}
0<\epsilon<\pi^{-\frac{1}{2}}.
\end{equation}
which gives a more rigorous meaning to the 
$\epsilon$ small assumption. We then compute the differences
\begin{equation}	
	\begin{aligned}
		\Delta x=\max_{t\in[0,1/\epsilon]}(|x-x_{ap}|)\\
		\Delta T=\max_{t\in[0,1/\epsilon]}(|T-T_{ap}|)\\		
	\end{aligned}\label{eq:Deltas}
\end{equation}
between the approximated and the numerical solutions 
of  eq.\eqref{eq:sysxT1} as functions of $\epsilon$ 
and one free initial condition while the remaining 
two are the equilibrium values (e.g. $x_0$ free, 
$\dot{x}_0=0$ and $T_0=0$). Figure \ref{fig:errorplotx} 
and Figure \ref{fig:errorplotT} show that, for a wide 
range of $\epsilon$ values and initial conditions, 
$\Delta x$ is $\BigO{\epsilon^3}$ while $\Delta T$ 
is $\BigO{\epsilon}$, which is consistent with the 
general properties of eq.\eqref{eq:solution} and 
eq.\eqref{eq:solution_T}.

\begin{figure}
\centering
\includegraphics[width=0.8\linewidth]{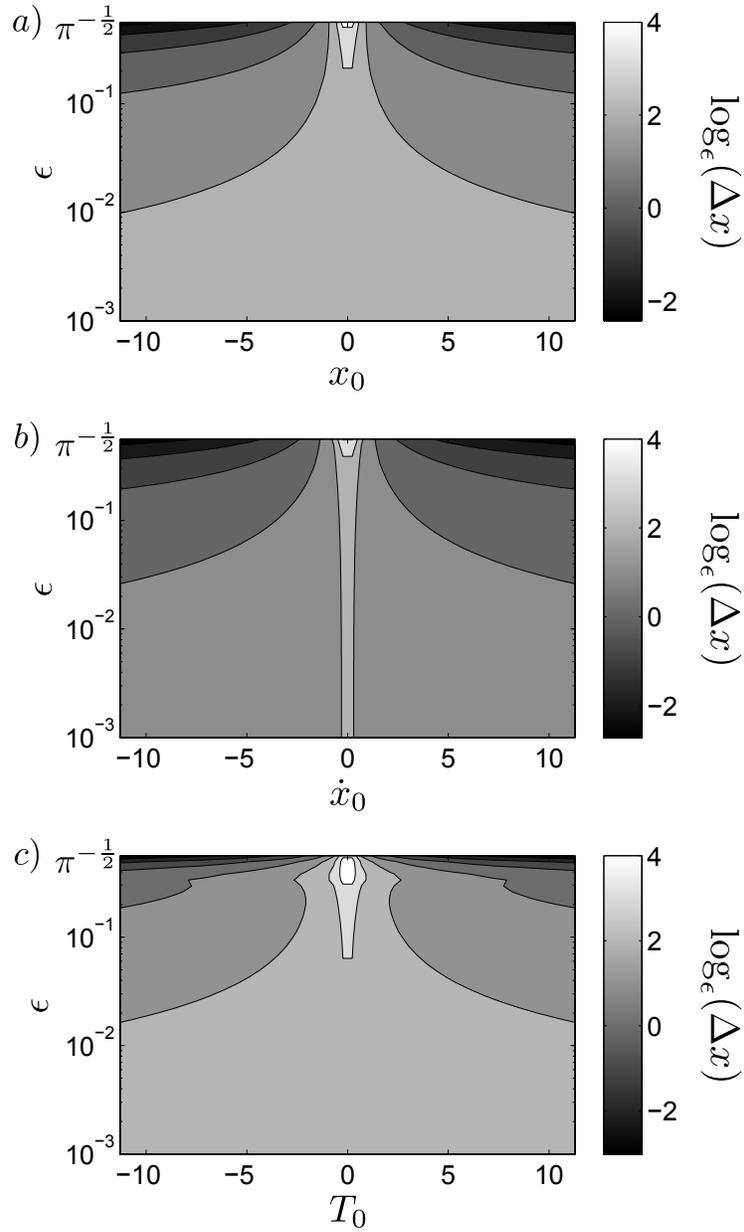}
\caption{Plots of $\log_{\epsilon} (\Delta {x})$ as 
a function of $\epsilon$, $x_0$, $\dot{x}_0$, $T_0$. 
Subplot a:  $x_0$ is varied with $\dot{x}_0=0$ and 
$T_0=0$. Subplot b:  $\dot{x}_0$ is varied with 
$x_0=0$ and $T_0=0$. Subplot c:  $T_0$ is varied 
with $x_0=0$ and $\dot{x}_0=0$. }\label{fig:errorplotx}
\end{figure}
\begin{figure}
\centering
\includegraphics[width=0.8\linewidth]{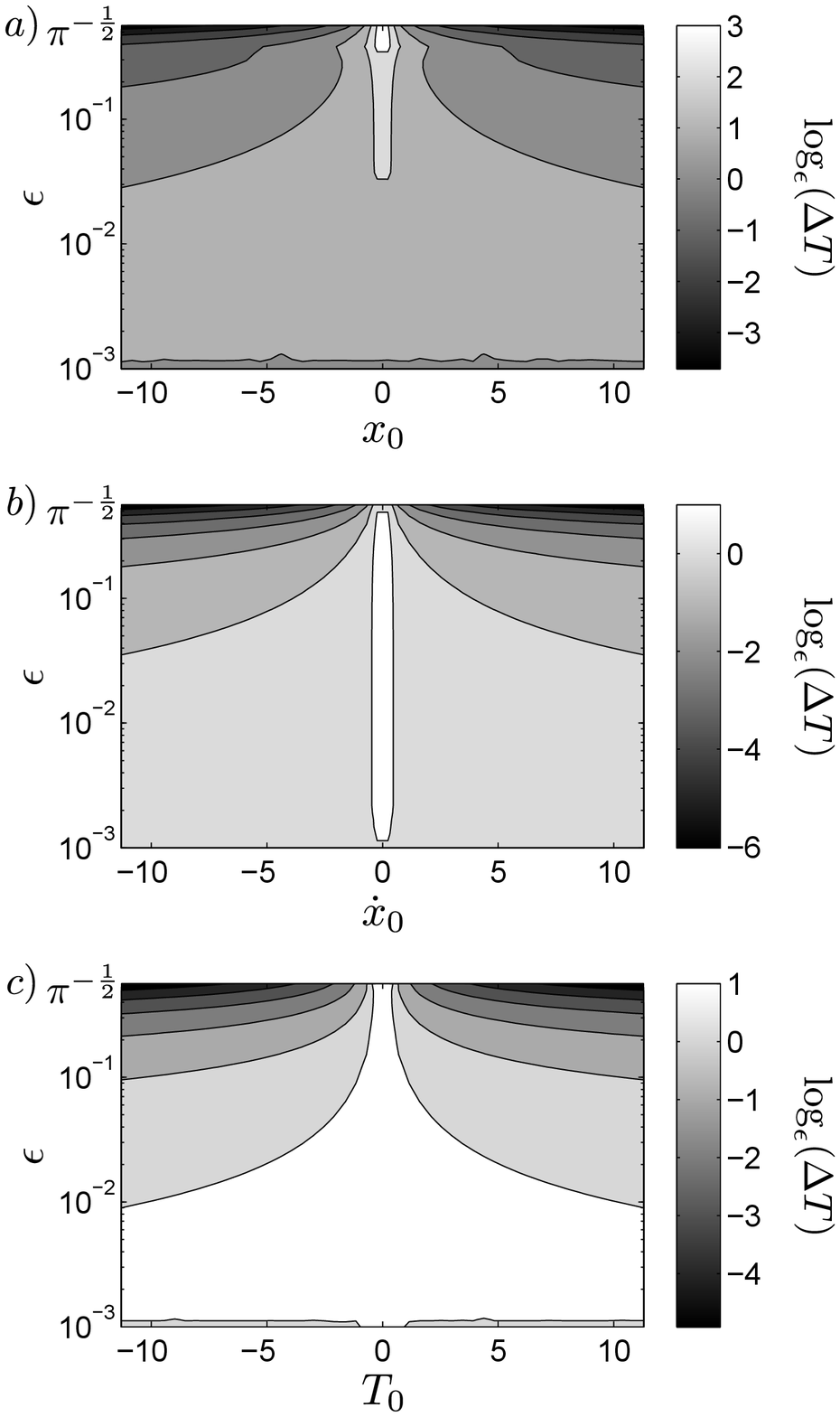}
\caption{Plots of $\log_{\epsilon} (\Delta {T})$ as 
a function of $\epsilon$, $x_0$, $\dot{x}_0$, $T_0$. 
Subplot a:  $x_0$ is varied with $\dot{x}_0=0$ and 
$T_0=0$. Subplot b:  $\dot{x}_0$ is varied with 
$x_0=0$ and $T_0=0$. Subplot c:  $T_0$ is varied 
with $x_0=0$ and $\dot{x}_0=0$. }\label{fig:errorplotT}
\end{figure}

We now compute the heat produced while the system 
relaxes to equilibrium. Substituting 
eq.\eqref{eq:autonomous_force} and $T_b=1$ in 
eq.\eqref{eq:Energetics_adim_Q}, gives
\begin{equation}
		Q^{rel}(0,t)=x(0)-x(t)+\frac{\dot{x}^2(0)-\dot{x}^2(t)}{2}+\frac{T(0)-T(t)}{2}
	\label{eq:heat_autonomous}
\end{equation}
Using now that $x(t)= x_{ap}(t)+\BigO{\epsilon^3}$ and that 
$T(t)= T_{ap}(t)+\BigO{\epsilon}$, we obtain the approximate 
expression for the heat
\begin{equation}
	\begin{aligned}
		Q_{ap}^{rel}(0,t)=&(1+x_0-x_{ap}(t))+\frac{\dot{x}_0^2-\dot{x}_{ap}^2(t)}{2}+\frac{1+T0-T_{ap}(t)}{2}+\BigO{\epsilon},
	\end{aligned}
	\label{eq:Q_relaxation}
\end{equation}
where we evaluated $x(0)=x_{ap}(0)$ and $T(0)=T_{ap}(0)$ 
instead of using eq.\eqref{eq:autonomous_BC}. 
This expression describes the heat exchanged with the 
reservoir as a relaxation process takes place and is fully 
analytic: numerical evaluation becomes a trivial 
task, while using it for some formal calculations  
is likely to allow results to be expressed in a closed form. 

\section{Compression in a finite time}
The second example we consider is the case in which 
the system, initially at equilibrium, undergoes a 
linear increase of the external force over a finite 
time and then relaxes to the new equilibrium condition. This encompasses all the isothermal compressions occurring in a finite time.
W.l.o.g. this is modeled by taking
\begin{equation}\label{eq:linearforce_protocol}
F=\begin{cases}
	1 &\text{ for } t<0\\
	1+f a \epsilon t &\text{ for } 0\leq t \leq\dfrac{1}{a\epsilon}\\
	1+f &\text{ for } t>\dfrac{1}{a\epsilon}.
\end{cases}
\end{equation}
with initial conditions
\begin{equation}\label{eq:linearforce_BC}
	\begin{aligned}
		x(0)&=1\\
		\dot{x}(0)&=0\\
		T(0)&=1.
\end{aligned}
\end{equation}
The additional parameters appearing in 
eq.\eqref{eq:linearforce_protocol} are the amount 
of force $f$ by which $F$ is increased and the 
constant $a$ which fixes the time-span $1/(a\epsilon)$.
Substituting eq.\eqref{eq:linearforce_protocol} in 
eq.\eqref{eq:solution} and then imposing that 
eq.\eqref{eq:linearforce_BC} must hold, we obtain after 
a long but straightforward calculation that, 
for $0\leq t\leq 1/(a\epsilon)$,
\begin{equation}\label{eq:linear_solution}
	\begin{aligned}
		x_{ap}(t)=&\frac{1}{1+f a \epsilon t}
		+\epsilon^2\bigg( -\frac{2a^2f^2}{(1+f a \epsilon t)^5}+\sqrt{\tfrac{2}{\pi}} \frac{af (\pi			+2)}{(1+f a \epsilon t)^3)}\bigg),
	\end{aligned}
\end{equation}
while for $t\geq 1/(a\epsilon)$, the system relaxes 
and $x_{ap}(t)$ is given by an adjusted version 
of eq.\eqref{eq:autonomous_solution} such that 
${x(\infty)=1/(f+1)}$, $\dot{x}(\infty)=0$ and $T(\infty)=1$.
We stress out that eq.\eqref{eq:linear_solution} is valid up to order $\BigO{\epsilon^4}$ because it satisfies equilibrium boundary conditions. Consequently, the corresponding $T_{ap}$ is valid up to $\BigO{\epsilon^3}$.

At this point we can compute eq.\eqref{eq:Deltas} 
as functions of $\epsilon$, $a$ and $f$ to 
investigate the quality of eq.\eqref{eq:linear_solution}. 
Since this parametric study does not yield results 
strikingly different from the ones we obtained for the 
relaxation case, we rather test the goodness of 
$x_{ap}$ by looking at the heat produced during the 
gas compression. If we neglect heat exchanges at 
intermediate times,  the net heat produced by this 
thermodynamic transformation is obtained by substitution 
of $T_{br}=1$ and eq.\eqref{eq:linearforce_protocol} 
into eq.\eqref{eq:Energetics_adim_Q}. After some 
simple calculations we obtain
\begin{equation}
	\begin{aligned}
		Q^{lin}(a)=&Q^{lin}(0,\infty,a)\\
		=&x(0)-1+\frac{\dot{x}^2(0)}{2}+\frac{T(0)-1}{2}
		+f\ a\ \epsilon \int_{0}^{\frac{1}{a\epsilon}}x(t) \mathrm{d} t,
	\end{aligned}
\end{equation}
where we dropped $t_i=0$ and $t_f=\infty$ for 
compactness and the new $a$ dependence is to 
remind that the compression occurs for 
$t\in[0,1/(a\epsilon)]$.
Substituting here eq.\eqref{eq:linear_solution} 
gives the approximated expression for the 
heat\footnote{As in the previous case, $x(0)=x_{ap}(0)$ and $T(0)=T_{ap}(0)$}
\begin{equation}
	\begin{aligned}
		Q_{ap}^{lin}(a)=&\ln(1+f) -\frac{2  (f  a \epsilon)^2 \left[(1+f)^2-\tfrac{1}{2}\right]\left[(1+f)+\tfrac{1}{2}					\right]}{(1+f)^4}\\
		&+\frac{2f a \epsilon^2 \sqrt{\tfrac{2}{\pi}} \left[(\pi+\tfrac{3}{2})(1+f)^2-\tfrac{\pi+2}{4}\right] }{(1+f)^2}+\BigO{\epsilon^3}
	\end{aligned}\label{eq:Q_linear}
\end{equation}
This formula is our main thermodynamic result: it 
estimates the heat produced by a perfect gas under 
the action of a finite-time compression. One interesting property of eq.\eqref{eq:Q_linear} is that in the limit of quasi-static transformations, i.e. $a\to 0$,
\begin{equation}
\lim_{a\to0} Q_{ap}^{lin}(a)=\ln(1+f).
\end{equation}
This is exactly the value prescribed by Clausius theorem. As a consequence, the remaining terms of eq.\eqref{eq:Q_linear} are contributions coming to the fact that the system is driven in a finite time. To have physical meaning, such contributions must be positive. This is a non-trivial requirement. However, the multiple scales method can be applied only if 
\begin{equation}
	\begin{aligned}
	f>&0\\
	a\lesssim&\frac{1}{f}.
	\end{aligned}\label{eq:validity_multiplescale}
\end{equation}
and, within these constrains, such positivity requirement is always satisfied.

We note that eq.\eqref{eq:validity_multiplescale} is not the only validity constrain because 
the worst protocol with 
the functional form of eq.\eqref{eq:linearforce_protocol} 
is the one where the compression is 
instantaneous. This corresponds to take the limit $a\rightarrow\infty$ in eq. \eqref{eq:Energetics_adim_Q} with eq.\eqref{eq:linearforce_BC} and yields the upper bound 
$Q^{lin}(a)\leq f$.  Therefore we must have that ${Q^{lin}(a)\in[\ln(1+f),f]}$. Since eq.\eqref{eq:Q_linear} has 
an evaluation error of the order of $\epsilon^2$, 
if the inequality
\begin{equation}\label{eq:validity_errorrange}
	f-\ln(1+f)\gg\epsilon^2
\end{equation}
is not satisfied, the estimation error on 
eq.\eqref{eq:Q_linear} is bigger than the energy 
region we want to investigate. Any result obtained 
with $Q^{lin}_{ap}(a)$ is then of no practical use. We 
therefore conclude that the validity region of 
eq.\eqref{eq:Q_linear} is given by 
eq.\eqref{eq:validity_multiplescale} and 
eq.\eqref{eq:validity_errorrange}. 

We conclude this section with a numerical study 
of eq.\eqref{eq:Q_linear}: we compute the difference 
\begin{equation}
	\Delta Q =|Q^{lin}(a)-Q_{ap}^{lin}(a)|\\
\end{equation} 
as functions of $a$ and $f$ for a given $\epsilon$ 
value. Figure \ref{fig:errorplotQ} and \ref{fig:errorplotQ_2} 
show the result obtained for $\epsilon=0.1$ and $\epsilon=0.01$. 
From the picture it clearly appears that $\Delta Q$ 
is $\BigO{\epsilon^3}$ within the constraints defined 
by eq.\eqref{eq:validity_multiplescale} and 
eq.\eqref{eq:validity_errorrange}. This 
proves that eq.\eqref{eq:Q_linear} is a good 
analytical expression of the heat produced during 
a finite time compression of a perfect gas.

\begin{figure}
\centering
\includegraphics[width=0.8\linewidth]{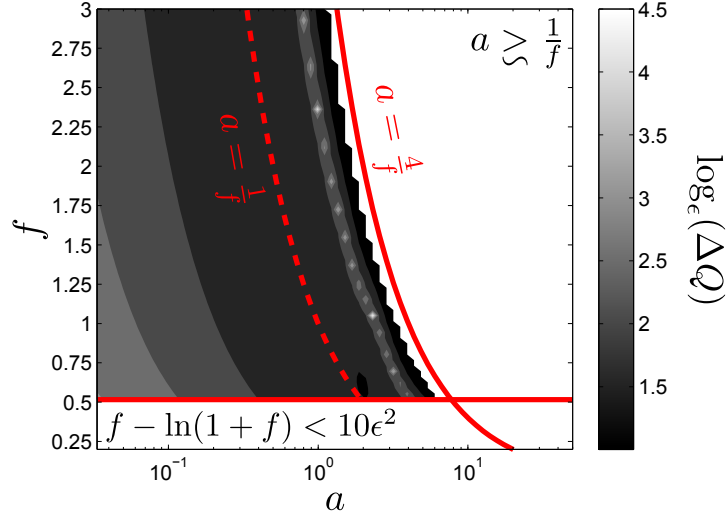}
\caption{Plot of $\log_{\epsilon}(\Delta Q)$ as 
a function of $f$ and $a$ for $\epsilon=0.1$. 
The region  $a\geq 4/f$ is left blank because 
eq.\eqref{eq:Q_linear} is defined only for 
$a\lesssim 1/f$ . The region $f-\ln(1+f)<10\epsilon^2$ 
is left blank because the estimation error on 
eq.\eqref{eq:Q_linear} makes any consideration 
on the produced heat meaningless if 
$f-\ln(1+f)<10\epsilon^2$ is not satisfied.}\label{fig:errorplotQ}
\end{figure}
\begin{figure}
\centering
\includegraphics[width=0.8\linewidth]{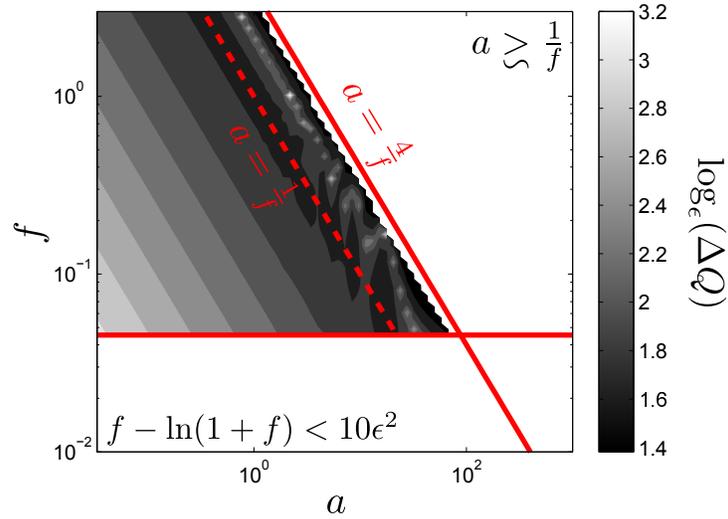}
\caption{Same as in Figure \ref{fig:errorplotQ}
but for $\epsilon=0.01$.}\label{fig:errorplotQ_2}
\end{figure}

\section{Conclusions and future developments}
In this paper we focused on the LDGPE derived 
in \cite{vulpiani1} to describe the dynamical 
evolution of a gas enclosed by a piston and in 
contact with a heat reservoir. In particular, 
we showed that the LDGPE have an approximated 
analytic solution when the temperature of the 
reservoir is fixed and the force applied on the 
piston varies according to a general force 
protocol. To derive such result we used the 
multiple scales expansion method. Although this 
is valid only within some constraints, we 
can use the approximated solutions to describe 
the thermodynamics of fundamental nonequilibrium 
processes. Our main result is that
 we are now able to compute 
in a closed form the heat produced when the 
gas-piston system 1) relaxes toward equilibrium, 
and 2) undergoes an isothermal compression in a 
finite time. As the derivation of analytic heat 
relations through the multiple scales method turned 
out to be quite straightforward, we believe that 
this perturbative technique could be useful
in understanding finite time thermodynamics. 

An issue that deserves futures 
investigations is the following: we already stated 
that eq.\eqref{eq:Q_linear} is valid when  
$a\lesssim1/f$. Nonetheless, it's clear that an 
instantaneous compression ($a=\infty$) produces 
heat only because the system relaxes toward the 
new equilibrium condition. It is not hard to 
prove by means of eq.\eqref{eq:Q_relaxation} that 
$Q^{lin}(\infty)=Q^{rel}_{ap}(0,\infty)=f$. We 
thus have that eq.\eqref{eq:solution} allows to 
simultaneously describe the heat produced by 
an isothermal compression for both $a\lesssim 1/f$ 
and $a\gg1/f$. This gives us an hint that there 
could be a way to access the missing $a$ region 
by means of some well aimed multiple scales 
expansion. One other issue that can be treated with the multiple scales method is the study of the resonance of eq.\eqref{eq:sysxT1}. Since the system has a unitary characteristic frequency, this problem can be efficiently addressed with the multiple scales method \cite{KevorkianCole,Holmes,BenderOrszag} by choosing 
\begin{equation}
F(t)= 1+\epsilon\sin\left((1+\epsilon)t\right),
\end{equation}
which is small in amplitude but not slow anymore. 
We also note that our results are a direct consequence 
of linearity of the LDGPE. However we are interested to derive a  
multiple scales expansion also for the DGPE.
 It is clear to the reader that this problem is far much difficult from the linear one because 1) the system does not reduce to a single equation and 2) $x$ and $T$ need different scalings to avoid singularities. 

We conclude by noting that if we restrict ourselves to dynamical equilibrium solutions we proved that it's possible to compute the heat exchanged when also the reservoir temperature is slowly varying over time \cite{ChiuchiuGubbiotti02}. The price to pay is that we lose the characterization of transient behaviors. However, those can be accessed by a full multiple scales method where also $T_b=T_b(\epsilon t)$.

\section*{Acknowledgments}
We would like to thank L. Gammaitoni, A. Vulpiani 
and L. Cerino for the useful discussion on 
the topic. DC is supported by the 
European union (FPVII(2007-2013) under G.A. n.318287 LANDAUER).
GG is supported by  INFN  IS-CSN4 \emph{Mathematical Methods of
Nonlinear Physics}.

\bibliographystyle{plain}
\bibliography{references_davide}

\begin{thebibliography}{10}

\bibitem{misprint}
private comunication with one of the authors.

\bibitem{balescu}
R.~Balescu.
\newblock {\em Equilibrium and Nonequilibrium Statistical Mechanics}.
\newblock Wiley-Blackwell, 1975.

\bibitem{BenderOrszag}
C.~M. Bender and S.~A. Orszag.
\newblock {\em Advanced mathematical methods for scientists and engineers}.
\newblock McGraw-Hill, 1978.

\bibitem{Calogero1991}
F.~Calogero.
\newblock Why are certain nonlinear pdes both widely applicable and integrable?
\newblock In V.E. Zakharov, editor, {\em What is integrability?} Springer,
  Berlin-Heidelberg, 1991.

\bibitem{CalogeroEchkhaus1987}
F.~Calogero and W.~Echkhaus.
\newblock Nonlinear evolution equations, rescalings, model pdes and their
  integrability, i.
\newblock {\em Inv. Probl.}, 3:229--262, 1987.

\bibitem{CalogeroEchkhaus1988}
F.~Calogero and W.~Echkhaus.
\newblock Nonlinear evolution equations, rescalings, model pdes and their
  integrability, ii.
\newblock {\em Inv. Probl.}, 4:11--13, 1988.

\bibitem{campisi1}
M.~Campisi, P.~H\"anggi, and P.~Talkner.
\newblock \textit{Colloquium} : Quantum fluctuation relations: Foundations and
  applications.
\newblock {\em Rev. Mod. Phys.}, 83:771--791, 2011.

\bibitem{campisi2}
M.~Campisi, J.~Pekola, and R.~Fazio.
\newblock Nonequilibrium fluctuations in quantum heat engines: theory, example,
  and possible solid state experiments.
\newblock {\em New J. of Phys.}, 17(3):035012, 2015.

\bibitem{vulpiani3}
M.~Cencini, L.~Palatella, S.~Pigolotti, and A.~Vulpiani.
\newblock Macroscopic equations for the adiabatic piston.
\newblock {\em Phys. Rev. E}, 76:051103, 2007.

\bibitem{cerino}
L.~Cerino, G.~Gradenigo, A.~Sarracino, D.~Villamaina, and A.~Vulpiani.
\newblock Fluctuations in partitioning systems with few degrees of freedom.
\newblock {\em Phys. Rev. E}, 89:042105, Apr 2014.

\bibitem{vulpiani1}
L.~Cerino, A.~Puglisi, and A.~Vulpiani.
\newblock Kinetic model for the finite-time thermodynamics of small heat
  engines.
\newblock {\em Phys. Rev. E}, 91:032128, 2015.

\bibitem{ChiuchiuGubbiotti02}
D~Chiuchi\`{u} and G.~Gubbiotti.
\newblock Nonequilibrium thermodynamics of a slow motor, 2016.
\newblock (in preparation).

\bibitem{ColeKevorkian1963}
J.~D. Cole and J.~Kevorkian.
\newblock Uniformly valid asymptotic approximations for certain differential
  equations.
\newblock In J.~P. LaSalle and S.~Lefschetz, editors, {\em Nonlinear
  differential equations and Nonlinear Mechanics}. Academic Press, New York,
  1963.

\bibitem{crooks}
G.~E. Crooks.
\newblock Nonequilibrium measurements of free energy differences for
  microscopically reversible {Markovian} systems.
\newblock {\em J. of Stat. Phy.}, 90:1481--1487.

\bibitem{degrootmazur}
S.~R. de~Groot and P.~Mazur.
\newblock {\em Non-equilibrium thermodynamics}.
\newblock North-Holland Publishing Company, 1962.

\bibitem{espositovandenbroeck}
M.~Esposito and C.~Van den Broeck.
\newblock Second law and {Landauer} principle far from equilibrium.
\newblock {\em EPL}, 95(4):40004, 2011.

\bibitem{feynman}
R.~P. Feynman.
\newblock {\em The Feynman Lectures on Physics}, volume~1.
\newblock Addison Wesley Longman, 1970.

\bibitem{gruber}
G.~Gruber and A.~Lesne.
\newblock {\em Encyclopedia of Mathematical Physics (Lemma: Adiabatic piston)},
  volume~1.
\newblock Elsevier, 2006.

\bibitem{Holmes}
M.~H. Holmes.
\newblock {\em Introduction to Perturbation Methods}.
\newblock Springer, 2013.

\bibitem{jarzynski}
C.~Jarzynsk.
\newblock Nonequilibrium equality for free energy differences.
\newblock {\em Phys. Rev. Lett.}, 78:2690--2693, 1997.

\bibitem{jarzynski2}
C.~Jarzynski.
\newblock Comparison of far-from-equilibrium work relations.
\newblock {\em C. R. Physique}, 8(5–6):495 -- 506, 2007.

\bibitem{KevorkianCole}
J.~Kevorkian and J.~D. Cole.
\newblock {\em Multiple scale and singular perturbation methods}.
\newblock Springer-Verlag, 1996.

\bibitem{Kuzmak1959}
G.E. Kuzmak.
\newblock Asymptotic solutions of nonlinear second order differential equations
  with variable coefficients.
\newblock {\em J. Appl. Math. Mech. (PMM)}, 23:730--744, 1959.

\bibitem{lieb}
E.~H. Lieb.
\newblock Some problems in statistical mechanics that i would like to see
  solved.
\newblock {\em Phys. A}, 263(1–4):491 -- 499, 1999.

\bibitem{Lindstedt1882}
A.~Lindstedt.
\newblock \"uber die integration einer f\"ur die st\"orungstheorie wichtigen
  differentialgleichung.
\newblock {\em Astron. Nachr.}, 103:211--220, 1882.

\bibitem{Nayfeh}
A.~H. Nayfeh.
\newblock {\em Perturbation Methods}.
\newblock John Wiley and Sons, 1973.

\bibitem{Poincare1886}
H.~Poincar\'e.
\newblock Sur les int\'egrales irr\'eguli\`eres des \'equations lin\'eaires.
\newblock {\em Acta Math.}, 8:295--344, 1886.

\bibitem{sano}
T.~G. {Sano} and H.~{Hayakawa}.
\newblock {Efficiency at maximum power output for a passive engine}.
\newblock Preprint, arXiv1412.4468.

\bibitem{seifert}
U.~Seifert.
\newblock Stochastic thermodynamics, fluctuation theorems and molecular
  machines.
\newblock {\em Rep. on Prog. in Phys.}, 75(12):126001, 2012.

\bibitem{sekimoto}
K.~Sekimoto.
\newblock {\em Stochastic Energetics}.
\newblock Springer-Verlag Berlin Heidelberg, 2010.

\bibitem{tehver}
R.~Tehver, F.~Toigo, J.~Koplik, and J.~R. Banavar.
\newblock Thermal walls in computer simulations.
\newblock {\em Phys. Rev. E}, 57:R17--R20, 1998.

\bibitem{ZakharovKuznetsov1986}
V.E. Zakharov and E.A. Kuznetsov.
\newblock Multi-scale expansions in the theory of systems integrable by the
  inverse scattering transform.
\newblock {\em Phys. D:}, 18:455--463, 1986.

\end{thebibliography}

\end{document}